\input harvmac
\input epsf

%
\let\includefigures=\iftrue
%
%
%
\newfam\black
\input rotate
\input epsf
\noblackbox
%
%
\includefigures
\message{If you do not have epsf.tex (to include figures),}
\message{change the option at the top of the tex file.}
\def\figin{\epsfcheck\figin}\def\figins{\epsfcheck\figins}
\def\epsfcheck{\ifx\epsfbox\UnDeFiNeD
\message{(NO epsf.tex, FIGURES WILL BE IGNORED)}
\gdef\figin##1{\vskip2in}\gdef\figins##1{\hskip.5in}
\else\message{(FIGURES WILL BE INCLUDED)}%
\gdef\figin##1{##1}\gdef\figins##1{##1}\fi}
\def\DefWarn#1{}
\def\N{{\cal N}}
\def\figinsert{\goodbreak\midinsert}
\def\ifig#1#2#3{\DefWarn#1\xdef#1{fig.~\the\figno}
\writedef{#1\leftbracket fig.\noexpand~\the\figno}%
\figinsert\figin{\centerline{#3}}\medskip\centerline{\vbox{\baselineskip12pt
\advance\hsize by -1truein\noindent\footnotefont{\bf
Fig.~\the\figno:} #2}}
\bigskip\endinsert\global\advance\figno by1}
\else
\def\ifig#1#2#3{\xdef#1{fig.~\the\figno}
\writedef{#1\leftbracket fig.\noexpand~\the\figno}%
\global\advance\figno by1} \fi

\def\tilde{\widetilde}

\def\yboxit#1#2{\vbox{\hrule height #1 \hbox{\vrule width #1
\vbox{#2}\vrule width #1 }\hrule height #1 }}
\def\fillbox#1{\hbox to #1{\vbox to #1{\vfil}\hfil}}
\def\ybox{{\lower 1.3pt \yboxit{0.4pt}{\fillbox{8pt}}\hskip-0.2pt}}

\def\rightarrowbox#1#2{
  \setbox1=\hbox{\kern#1{${ #2}$}\kern#1}
  \,\vbox{\offinterlineskip\hbox to\wd1{\hfil\copy1\hfil}
    \kern 3pt\hbox to\wd1{\rightarrowfill}}}

\def\Tr{{{\rm Tr~ }}}

\def\tilde{\widetilde}

\def\II{\relax{I\kern-.10em I}}

\def\bar{\overline}

\def\IZ{\relax\ifmmode\mathchoice
{\hbox{\cmss Z\kern-.4em Z}}{\hbox{\cmss Z\kern-.4em Z}}
{\lower.9pt\hbox{\cmsss Z\kern-.4em Z}} {\lower1.2pt\hbox{\cmsss
Z\kern-.4em Z}}\else{\cmss Z\kern-.4em Z}\fi}
\def\IB{\relax{\rm I\kern-.18em B}}
\def\IC{{\relax\hbox{$\inbar\kern-.3em{\rm C}$}}}
\def\ID{\relax{\rm I\kern-.18em D}}
\def\IE{\relax{\rm I\kern-.18em E}}
\def\IF{\relax{\rm I\kern-.18em F}}
\def\IG{\relax\hbox{$\inbar\kern-.3em{\rm G}$}}
\def\IGa{\relax\hbox{${\rm I}\kern-.18em\Gamma$}}
\def\IH{\relax{\rm I\kern-.18em H}}
\def\II{\relax{\rm I\kern-.18em I}}
\def\IK{\relax{\rm I\kern-.18em K}}
\def\IN{\relax{\rm I\kern-.18em N}}
\def\IP{\relax{\rm I\kern-.18em P}}

%
\def\inbar{\,\vrule height1.5ex width.4pt depth0pt}

\font\cmss=cmss10 \font\cmsss=cmss10 at 7pt
\def\IR{\relax{\rm I\kern-.18em R}}

\def\lp10{l_P^{10}}
\def\lp11{l_P^{11}}
\def\R11{R_{11}}

\newbox\tmpbox\setbox\tmpbox\hbox{\abstractfont
}
 \Title{\vbox{\baselineskip12pt\hbox to\wd\tmpbox{\hss
 hep-th/0409245} }}
 {\vbox{\centerline{Gauge Theory Amplitudes In Twistor Space}
 \bigskip
 \centerline{And Holomorphic Anomaly}
 }}
\smallskip
\centerline{Freddy Cachazo$^*$, Peter Svr\v cek$^\#$, and Edward
Witten$^*$} \bigskip \centerline{\it $^*$ School of Natural
Sciences, Institute for Advanced Study, Princeton NJ 08540
USA}\bigskip \centerline{\it $^\#$ Department of Physics, Joseph
Henry Laboratories, Princeton NJ 08540 USA}
\bigskip
\vskip 1cm \noindent

\input amssym.tex
We show that, in analyzing differential equations obeyed by
one-loop gauge theory amplitudes, one must take into account a
certain holomorphic anomaly.  When this is done, the results are
consistent with the simplest twistor-space picture of the
available one-loop amplitudes.

\Date{September 2004}

\lref\BernZX{ Z.~Bern, L.~J.~Dixon, D.~C.~Dunbar and
D.~A.~Kosower, ``One Loop $N$-Point Gauge Theory Amplitudes,
Unitarity And Collinear Limits,'' Nucl.\ Phys.\ B {\bf 425}, 217
(1994), hep-ph/9403226.
}

\lref\BernCG{ Z.~Bern, L.~J.~Dixon, D.~C.~Dunbar and
D.~A.~Kosower, ``Fusing Gauge Theory Tree Amplitudes into Loop
Amplitudes,'' Nucl.\ Phys.\ B {\bf 435}, 59 (1995),
hep-ph/9409265.
}

\lref\WittenNN{ E.~Witten, ``Perturbative Gauge Theory as a String
Theory in Twistor Space,'' hep-th/0312171.
}

\lref\CachazoKJ{ F.~Cachazo, P.~Svrcek and E.~Witten, ``MHV
Vertices and Tree Amplitudes in Gauge Theory,'' hep-th/0403047.
}

\lref\berkwitten{N. Berkovits and E. Witten,  ``Conformal
Supergravity In Twistor-String Theory,'' hep-th/0406051.}

\lref\penrose{R. Penrose, ``Twistor Algebra,'' J. Math. Phys. {\bf
8} (1967) 345.}

\lref\berends{F. A. Berends, W. T. Giele and H. Kuijf, ``On
Relations Between Multi-Gluon And Multi-Graviton Scattering,"
Phys. Lett {\bf B211} (1988) 91.}

\lref\berendsgluon{F. A. Gerends, W. T. Giele and H. Kuijf,
``Exact and Approximate Expressions for Multigluon Scattering,"
Nucl. Phys. {\bf B333} (1990) 120.}

\lref\bernplusa{Z. Bern, L. Dixon and D. A. Kosower, ``New QCD
Results From String Theory,'' in {\it Strings '93}, ed. M. B.
Halpern et. al. (World-Scientific, 1995), hep-th/9311026.}

\lref\bernplusb{Z. Bern, G. Chalmers, L. J. Dixon and D. A.
Kosower, ``One Loop $N$ Gluon Amplitudes with Maximal Helicity
Violation via Collinear Limits," Phys. Rev. Lett. {\bf 72} (1994)
2134.}

\lref\bernfive{Z. Bern, L. J. Dixon and D. A. Kosower, ``One Loop
Corrections to Five Gluon Amplitudes," Phys. Rev. Lett {\bf 70}
(1993) 2677.}

\lref\bernfourqcd{Z.Bern and  D. A. Kosower, "The Computation of
Loop Amplitudes in Gauge Theories," Nucl. Phys.  {\bf B379,}
(1992) 451.}

\lref\cremmerlag{E. Cremmer and B. Julia, ``The $N=8$ Supergravity
Theory. I. The Lagrangian," Phys. Lett.  {\bf B80} (1980) 48.}

\lref\cremmerso{E. Cremmer and B. Julia, ``The $SO(8)$
Supergravity," Nucl. Phys.  {\bf B159} (1979) 141.}

\lref\dewitt{B. DeWitt, "Quantum Theory of Gravity, III:
Applications of Covariant Theory," Phys. Rev. {\bf 162} (1967)
1239.}

\lref\dunbarn{D. C. Dunbar and P. S. Norridge, "Calculation of
Graviton Scattering Amplitudes Using String Based Methods," Nucl.
Phys. B {\bf 433,} 181 (1995), hep-th/9408014.}

\lref\ellissexton{R. K. Ellis and J. C. Sexton, "QCD Radiative
corrections to parton-parton scattering," Nucl. Phys.  {\bf B269}
(1986) 445.}

\lref\gravityloops{Z. Bern, L. Dixon, M. Perelstein, and J. S.
Rozowsky, ``Multi-Leg One-Loop Gravity Amplitudes from Gauge
Theory,"  hep-th/9811140.}

\lref\kunsztqcd{Z. Kunszt, A. Singer and Z. Tr\'{o}cs\'{a}nyi,
``One-loop Helicity Amplitudes For All $2\rightarrow2$ Processes
in QCD and ${\cal N}=1$ Supersymmetric Yang-Mills Theory,'' Nucl.
Phys.  {\bf B411} (1994) 397, hep-th/9305239.}

\lref\mahlona{G. Mahlon, ``One Loop Multi-photon Helicity
Amplitudes,'' Phys. Rev.  {\bf D49} (1994) 2197, hep-th/9311213.}

\lref\mahlonb{G. Mahlon, ``Multi-gluon Helicity Amplitudes
Involving a Quark Loop,''  Phys. Rev.  {\bf D49} (1994) 4438,
hep-th/9312276.}

\lref\klt{H. Kawai, D. C. Lewellen and S.-H. H. Tye, ``A Relation
Between Tree Amplitudes of Closed and Open Strings," Nucl. Phys.
{B269} (1986) 1.}

\lref\pppmgr{Z. Bern, D. C. Dunbar and T. Shimada, ``String Based
Methods In Perturbative Gravity," Phys. Lett.  {\bf B312} (1993)
277, hep-th/9307001.}

\lref\GiombiIX{ S.~Giombi, R.~Ricci, D.~Robles-Llana and
D.~Trancanelli, ``A Note on Twistor Gravity Amplitudes,''
hep-th/0405086.
}

\lref\WuFB{ J.~B.~Wu and C.~J.~Zhu, ``MHV Vertices and Scattering
Amplitudes in Gauge Theory,'' hep-th/0406085.
}

\lref\Feynman{R.P. Feynman, Acta Phys. Pol. 24 (1963) 697, and in
{\it Magic Without Magic}, ed. J. R. Klauder (Freeman, New York,
1972), p. 355.}

\lref\Peskin{M.E. Peskin and D.V. Schroeder, {\it An Introduction
to Quantum Field Theory} (Addison-Wesley Pub. Co., 1995).}

\lref\parke{S. Parke and T. Taylor, ``An Amplitude For $N$ Gluon
Scattering,'' Phys. Rev. Lett. {\bf 56} (1986) 2459; F. A. Berends
and W. T. Giele, ``Recursive Calculations For Processes With $N$
Gluons,'' Nucl. Phys. {\bf B306} (1988) 759. }

\lref\BrandhuberYW{ A.~Brandhuber, B.~Spence and G.~Travaglini,
``One-Loop Gauge Theory Amplitudes In N = 4 Super Yang-Mills From
MHV Vertices,'' hep-th/0407214.
}

\lref\CachazoZB{ F.~Cachazo, P.~Svrcek and E.~Witten, ``Twistor
Space Structure Of One-Loop Amplitudes In Gauge Theory,''
hep-th/0406177.
}

\lref\RoiSpV{R.~Roiban, M.~Spradlin and A.~Volovich,
JHEP {\bf 0404}, 012 (2004) hep-th/0402016; R.~Roiban and
A.~Volovich, ``All Googly Amplitudes From The $B$-Model In Twistor
Space,'' hep-th/0402121; R.~Roiban, M.~Spradlin and A.~Volovich,
``On The Tree-Level S-Matrix Of Yang-Mills Theory,'' Phys.\ Rev.\
D {\bf 70}, 026009 (2004) hep-th/0403190.}

\lref\Gukovetal{S.~Gukov, L.~Motl and A.~Neitzke, ``Equivalence Of
Twistor Prescriptions For Super Yang-Mills,'' hep-th/0404085.}

\newsec{Introduction}

Perturbative scattering amplitudes of gluons in Yang-Mills theory
exhibit many remarkable properties, some of which \WittenNN\
apparently reflect the fact that they can be computed using a
string theory in which twistor space is the target space.

At tree level, the amplitudes can be constructed either from
connected diagrams in twistor space, as described most fully in
\RoiSpV, or from disconnected diagrams \CachazoKJ. In the latter
approach, one constructs the tree amplitudes from Feynman diagrams
in which the vertices are tree-level MHV (maximal helicity
violating) amplitudes, continued off-shell, and the propagators
are simple Feynman propagators $1/p^2$.  For a discussion of the
relation between these two approaches, see \Gukovetal.

At one-loop, the twistor space structure of various gauge theory
amplitudes was studied in \CachazoZB\ by studying the differential
equations that they obey.  The simple twistor space picture of
\WittenNN\ suggests that the one-loop MHV amplitudes of ${\cal
N}=4$ super Yang-Mills theory are a sum of terms corresponding to
the twistor space pictures of figures 1a and 1b -- all gluons
supported on a pair of lines (figure 1a) or on a degree two curve
of genus zero (figure 1b). However, the differential equations
appear to indicate the presence of a further contribution (figure
1c) in which all gluons but one are on a pair of lines,  while the
remaining gluon is coplanar with the two lines.\foot{The
differential equations seem to indicate that the degree two curve
of figure 1b degenerates to a pair of intersecting lines, as is
drawn in the figure.  This behavior is not entirely understood,
and needs to be interpreted with care, as explained in
\CachazoZB.} This is perplexing, because there has been no
proposal for a twistor-string mechanism to generate a contribution
with this structure.

\ifig\tomu{Shown here are twistor configurations that were found
in \CachazoZB\ to contribute to one-loop supersymmetric MHV
amplitudes. In (a), all gluons are inserted on a pair of possibly
disjoint lines connected by two twistor space propagators.  In
(b), all gluons are inserted on a degree two curve of genus zero
with a twistor space propagator; the curve is here drawn simply as
a pair of intersecting lines. (c) is just like (b) except that one
gluon is inserted not on the pair of intersecting lines but
somewhere else in the plane containing the two lines. In the
figures, dashed lines indicate twistor space propagators.}
{\epsfxsize=0.90\hsize\epsfbox{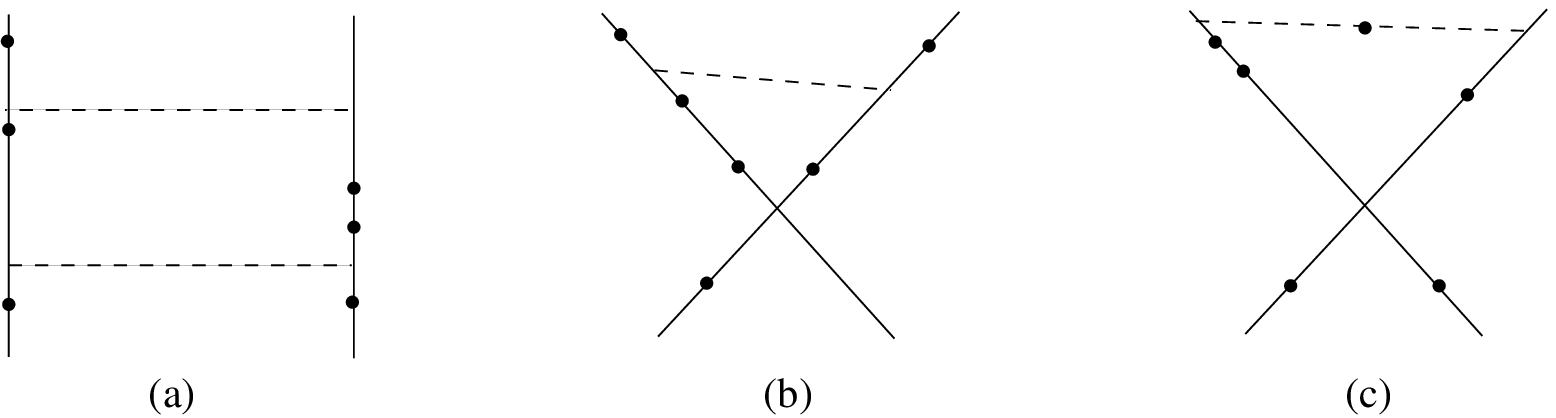}}

We have reconsidered these issues because of  a  new computation
of the one-loop $\N =4$ MHV amplitudes  \BrandhuberYW. In that
computation, the full amplitude is obtained  from a  one-loop
amplitude with two MHV vertices and $1/p^2$ propagators (figure
2); this is a direct one-loop generalization of the tree diagrams
considered in \CachazoKJ. Since each MHV vertex is supported on a
line in twistor space, this computation seems to make it manifest
that these one-loop MHV amplitudes are supported on a pair of
lines, the configuration of figure 1a. The configuration of figure
1b is also possible (when one propagator collapses) given that the
degree two curve in figure 1b is really a pair of intersecting
lines, as mentioned in the last footnote. But a configuration of
the type of figure 1c seems to be missing.

\ifig\lale{Diagram contributing to a one-loop ${\cal N}=4$ MHV
amplitude. Each disc represents a ``line," that is a $\Bbb{CP}^1$
in twistor space, which generates a tree-level MHV amplitude with
gluons attached. The loop amplitude is computed by connecting two
such MHV vertices via exchange of two gluons.}
{\epsfxsize=0.50\hsize\epsfbox{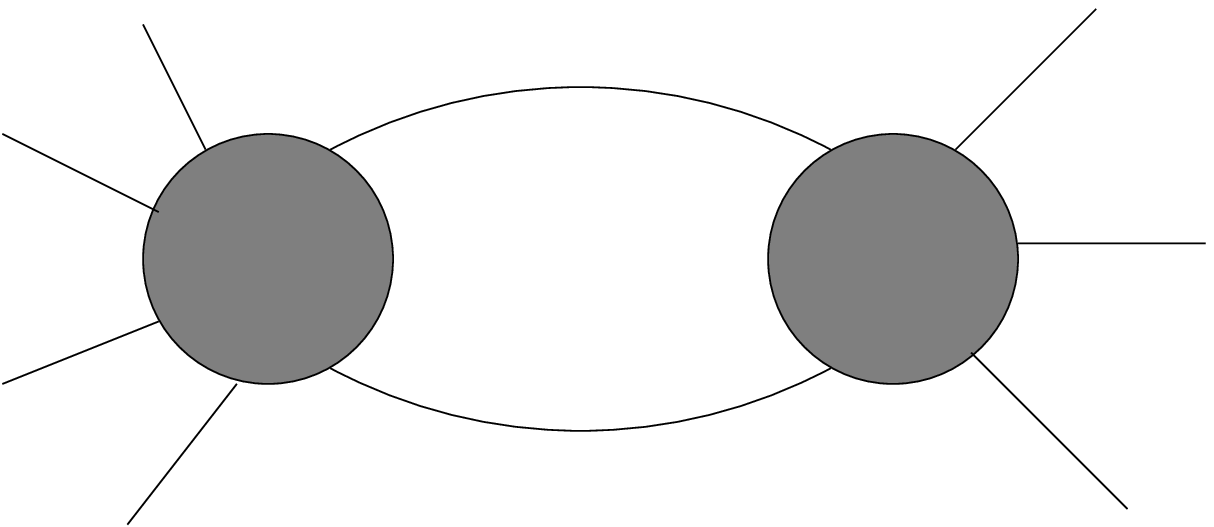}}

This problem can be posed more sharply if one considers the
imaginary part of the scattering amplitude.  Even without invoking
the full cut-constructibility of these one-loop amplitudes
\BernZX, one knows just from unitarity that for real momenta in
Lorentz signature, the imaginary part of the scattering amplitude
comes from a sum over on-shell intermediate states.  Thus, the
imaginary part of the amplitude can be obtained from the ``cut''
diagram of figure 3, where the ``cut'' propagators are on-shell
and the scattering amplitudes on the left and right are on-shell
tree level MHV amplitudes.  This seems to show that at least the
discontinuities (or on-shell imaginary parts) of the scattering
amplitudes must be supported on a pair of lines.  However, when we
investigate the differential equations obeyed by the imaginary
part of the scattering amplitudes, we find (as one would guess
from the fact that these amplitudes can be constructed from their
four-dimensional cuts) that the imaginary parts obey the same
differential equations as the full amplitudes; they do not obey
additional equations which would assert the absence of a
contribution of the type of figure 1c.

\ifig\mike{``Cut" diagram. Left and right tree-level MHV
amplitudes are on-shell. Internal lines represent the legs coming
from the ``cut" propagators.}
{\epsfxsize=0.50\hsize\epsfbox{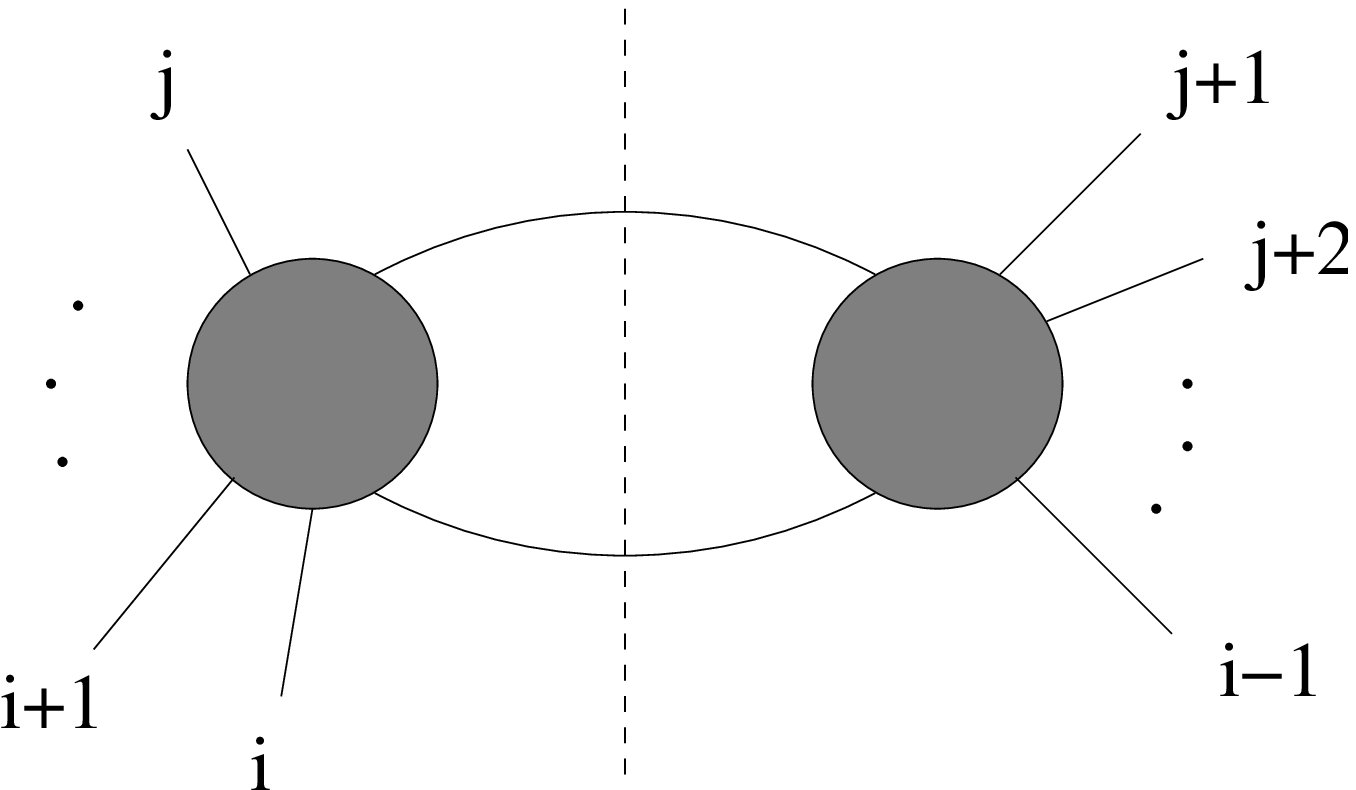}}

So we face an apparent contradiction, which we will resolve in the
present paper. The resolution turns out to be what one might
characterize as a holomorphic anomaly in the scattering
amplitudes, or in the differential equations that one uses to
analyze them. The ``cut'' diagram of figure 3 generates an
amplitude that seems to be manifestly supported on a pair of
lines, since the tree level MHV amplitude to the left or right of
the cut is supported on a line. However, our criterion for
asserting that an amplitude is supported on a pair of lines is
that it should be annihilated by certain differential operators;
when we act on the cut diagram with these operators, we get a
non-zero result because of a certain holomorphic anomaly. The
anomaly arises when one of the internal lines is collinear with an
adjacent external line; this can occur for generic on-shell
external momenta.

We think that the best summary of the facts is that the cut
amplitude is indeed supported on a pair of lines, and the subtlety
is in the use of the differential equations as a criterion for
investigating this.   The same should be true of the full
amplitude, since the amplitudes are cut-constructible. The
contributions of figure 1(c) found in \CachazoZB\ (in theories
with varying amounts of supersymmetry) are thus no longer
necessary, and the one-loop amplitudes, even with reduced
supersymmetry, may potentially be generated by simple
twistor-string theories.

In the next section, we explain the basic idea of the holomorphic
anomaly and show that it produces a contribution to the cut with
the structure of figure 1c.

\newsec{The Holomorphic Anomaly}

We will consider tree-level scattering of $n$ gluons, labeled by
$i=1,\dots,n$, with momenta $p_i^{a\dot
a}=\lambda_i^a\tilde\lambda_{i}^{\dot a}$.  We use spinor notation
with conventions described in \WittenNN.

We focus on MHV amplitudes in which two gluons, say $j$ and $k$,
have negative helicity, while the others have positive helicity.
We consider a subamplitude in which the gluons are in the cyclic
order $123\dots n$ and accordingly the group theory factor is
$\Tr\,T_1T_2\dots T_n$.  This subamplitude (with the group theory
factor and a momentum-conserving delta function
$-ig^{n-2}\delta^4(\sum_i\lambda_i^a\tilde\lambda_i^{\dot a})$
omitted) is \eqn\norfox{A(\lambda_i,\tilde\lambda_i)=\langle
\lambda_j,\lambda_k\rangle^4 \prod_{m=1}^n{1\over \langle
\lambda_m,\lambda_{m+1}\rangle}.} The fact that the amplitude is a
function only of $\lambda$ and not $\tilde\lambda$ is described by
saying that it is holomorphic.  The terminology reflects the fact
that for real momenta in Minkowski space, $\tilde\lambda$ is plus
or minus the complex conjugate of $\lambda$ (the sign
distinguishes initial and final particles).

The holomorphy was interpreted in \WittenNN\ to mean that the $n$
gluons in this scattering amplitude are supported on a line in
twistor space.  This collinearity was also expressed in terms of a
differential equation.  For this, we associate with a particle of
momentum $p^{a\dot a}=\lambda^a\tilde\lambda^{\dot a}$ its twistor
coordinates $Z^I=(\lambda^a,\mu^{\dot a})$, where $\mu_{\dot a}=
-i\partial/\partial\tilde\lambda^{\dot a}$.  The condition that
particles $i_1,i_2,$ and $i_3$ are collinear in twistor space is
that $\epsilon_{IJKL}Z_{i_1}^IZ_{i_2}^JZ_{i_3}^K=0$ for all $L$.
For example, if we take $L=\dot a$,  the condition is that
$\langle\lambda_{i_1},\lambda_{i_2}\rangle\mu_{i_3}+
\langle\lambda_{i_2},\lambda_{i_3}\rangle\mu_{i_1}
+\langle\lambda_{i_3},\lambda_{i_1}\rangle\mu_{i_2}=0$. Expressing
$\mu$ in terms of $\partial/\partial\tilde\lambda$, the
differential operator that should annihilate a scattering
amplitude that is supported on a line is \eqn\mumbo{F_{i_1i_2i_3}=
\langle\lambda_{i_1},\lambda_{i_2}\rangle{\partial\over\partial\tilde
\lambda_{i_3}}+ \langle\lambda_{i_2},\lambda_{i_3}\rangle
{\partial\over\partial\tilde \lambda_{i_1}}
+\langle\lambda_{i_3},\lambda_{i_1}\rangle
{\partial\over\partial\tilde \lambda_{i_2}}.} The amplitude
\norfox\ is a function of the $\lambda$'s only, so it appears to
be manifestly annihilated by $F_{i_1i_2i_3}$, for all $i_1,i_2,$
and $i_3$.

This is so for generic momenta, but there actually is a delta
function contribution when two adjacent gluons (in the cyclically
ordered chain $123\dots n$) become collinear. To see this, we
first rewrite $\tilde\lambda$ as $\bar\lambda$, as is appropriate
for real momenta.\foot{Even if one does not want the momenta to be
real in Lorentz signature, to carry out the one-loop integrals,
one needs an integration contour in which $\tilde\lambda$ is a
non-holomorphic function of $\lambda$; any such contour will lead
to a similar result.} The differential operator that should
annihilate the amplitudes is then \eqn\tumbo{F_{i_1i_2i_3}=
\langle\lambda_{i_1},\lambda_{i_2}\rangle{\partial\over\partial\bar
\lambda_{i_3}}+ \langle\lambda_{i_2},\lambda_{i_3}\rangle
{\partial\over\partial\bar \lambda_{i_1}}
+\langle\lambda_{i_3},\lambda_{i_1}\rangle
{\partial\over\partial\bar \lambda_{i_2}}.} It seems that this
operator does annihilate the MHV amplitude
$A(\lambda_i,\bar\lambda_i)$, because this amplitude is
holomorphic in the $\lambda_i$.  But we must be careful; this
function actually has poles, and these poles lead to delta
functions when we act with the differential operators.

The basic fact that gives rise to the delta functions is that, for
spinor variables $\lambda$ and $\lambda'$, \eqn\hutof{
d\bar\lambda{}^{\dot a}{\partial\over\partial\bar\lambda{}^{\dot
a}}{1\over
\langle\lambda,\lambda'\rangle}=2\pi\bar\delta(\langle\lambda,\lambda'\rangle).}
(Here, for a complex variable $z$, $\bar\delta(z)=-id\bar
z\delta^2(z)$.)   When this formula is inserted in the evaluation
of $F_{i_1i_2i_3}A$, we get not zero, but rather delta function
contributions when two adjacent gluons are collinear, that is,
when they obey  $\langle\lambda,\lambda'\rangle=0$. This does not
seem so important for tree diagrams, where for generic initial and
final momenta, this collinearity does not arise.  But the delta
functions are significant when the tree-level MHV vertices are
used to generate subamplitudes in a larger picture such as that of
figure 3, where we must integrate over momenta of internal gluons,
which may become collinear with one of the external gluons.   (The
integration region in computing the cut is compact, and there are
no singularities other than the collinear singularities we
consider momentarily, so these singularities are the only source
of an anomaly.)

In figure 3, for generic initial and final states, no two external
momenta are collinear, and  energy-momentum conservation does not
permit the two internal lines -- represented by ``cut''
propagators in the diagram -- to be collinear.  The important
case, therefore, is that an external gluon is collinear with one
of the internal gluons. Moreover, because of the form of the MHV
tree amplitude $A$, a pole only arises if this collinearity
involves an external and internal gluon that are adjacent in one
of the vertices. In figure 3, we  consider a configuration in
which external particles $i,i+1,\dots , j$ (for some $i$ and $j$)
couple to one MHV vertex and external particles
$j+1,j+2,\dots,i-1$ couple to the other vertex. Each vertex thus
contains a chain of external gluons as well as the two internal
gluons.  A pole in the MHV amplitude associated with the vertex,
leading to a delta function in the differential equations, arises
if one of the gluons at the end of one of the chains (gluon $i$ or
$j$ at one vertex, or gluon $j+1$ or $i-1$ at the other) is
collinear with the adjacent internal gluon.

Whatever the external momenta may be, energy-momentum conservation
allows any specified internal gluon to be collinear with any
specified initial or final state gluon. This is obvious in the
center of mass frame, where energy-momentum conservation fixes the
energy of the internal gluons (each has half the total center of
mass energy), while their directions of spatial motion are
opposite but otherwise arbitrary.  Hence, either internal gluon
may propagate in any specified direction.

Notice, however, that once we require one internal gluon to be
collinear with a given external gluon, the internal momenta in the
loop are completely determined. So, for generic external momenta,
we do not have the freedom to make {\it each} internal gluon
collinear with one of the external gluons.

{}From this discussion, it follows that if we act on the cut
amplitude $B$ of figure 3 with $F_{i_1i_2i_3}$, where gluons
$i_1$, $i_2$, and $i_3$ couple to the same MHV vertex, we will get
zero if none of the particles $i_1,i_2,$ or $i_3$ are at the end
of one of the chains.   If one of $i_1,i_2$, or $i_3$ is at the
end of a chain, the action of $F$ will give a delta function,
which then, upon integrating over the internal momenta, will give
a nonzero result. So for these choices of the gluons,
$F_{i_1i_2i_3}$ does not annihilate the amplitude.

Since $F_{i_1i_2i_3}B=0$, where $i_1,i_2$, and $i_3$ are disjoint
from the ends of the chains, it follows that all of the gluons at
a vertex that are not at an end of their respective chain are
supported on a line.  Thus, as there are two vertices, all gluons
except the endpoint gluons $i,j,j+1$, and $i-1$ are supported on a
union of two lines.  Moreover, all but one of the endpoint gluons
are supported on  the two lines. To prove this, we simply write
the integral over internal momenta in the cut amplitude as a sum
of integrals over subregions each of which only contains one pole.
This is possible for a reason observed earlier: for generic
external momenta, energy-momentum conservation does not allow two
distinct endpoint gluons, such as particles $i$ and $j$, to each
be collinear with an internal gluon.  Containing only one pole,
each subregion contributes to the amplitude a term in which at
most one gluon is not contained in the pair of lines.  So the full
amplitude is a sum of terms in each of which, from the standpoint
of the differential equations, all gluons but one are contained in
the union of two lines.

We can express this reasoning in terms of differential equations
by observing that the contribution to the cut from the diagram of
figure 3 is annihilated by suitable products of $F$'s, for example
\eqn\unin{F_{ii_1i_2}F_{jj_1j_2}B=0,} where $i,i_1,i_2$ and
likewise $j,j_1,j_2$ each couple to the same vertex.  This holds,
again, because anomalies involving the different $F$'s come from
disjoint regions of the integration over internal momenta.
Eqn.\unin\ is interpreted to mean that $B$ is a sum of terms
annihilated by one $F$ or the other, that is, a sum of terms in
which either particle $i$ or particle $j$ is contained in the
union of the two lines.

The holomorphic anomaly thus leads to precisely the situation that
actually was found in \CachazoZB: if one uses the differential
equations as a criterion for collinearity, then  $n-1$ of the $n$
gluons are contained in two chains of consecutive gluons, with
each chain being supported on a line in twistor space. The
remaining gluon is generically not contained in this union of
lines. We think, however, that it is most natural to describe the
cut amplitude of figure 3 as being supported on a pair of lines,
because in twistor-string theories we expect this kind of
amplitude to be generated by contributions that intuitively are
supported on a pair of lines. We prefer to interpret the
phenomenon we have found as a subtlety in the use of the
differential equations as a criterion for collinearity.

\bigskip
\bigskip
\centerline{\bf Acknowledgements}

Work of F. Cachazo was supported in part by the Martin A. and
Helen Chooljian Membership at the Institute for Advanced Study and
by DOE grant DE-FG02-90ER40542; that of P. Svrcek by NSF grants
PHY-9802484 and PHY-0243680; and that of E. Witten by NSF grant
PHY-0070928. Opinions and conclusions expressed here are those of
the authors and do not necessarily reflect the views of funding
agencies.

\listrefs

\end